\documentclass[conference]{IEEEtran}
\IEEEoverridecommandlockouts
\usepackage{cite}
\usepackage{amsmath,amssymb,amsfonts}

\newtheorem{proposition}{Proposition}

\newtheorem{property}{Property}
\newtheorem{definition}{Definition}
\usepackage{pifont}
\newcommand{\cmark}{\ding{51}}%
\newcommand{\xmark}{\ding{55}}%
\usepackage{graphicx}
\usepackage{subcaption}
\usepackage{mathtools}
\usepackage{algorithmic}
\usepackage{graphicx}
\usepackage{textcomp}
\usepackage{url}
\usepackage{bigints}
\usepackage{xcolor}

\usepackage{tcolorbox}

\usepackage{xcolor}
\def\BibTeX{{\rm B\kern-.05em{\sc i\kern-.025em b}\kern-.08em
    T\kern-.1667em\lower.7ex\hbox{E}\kern-.125emX}}
\begin{document}
\title{Towards faster settlement in HTLC-based Cross-Chain Atomic Swaps\\
\thanks{This work was supported by the European Research Council (ERC) under the European Union’s Horizon 2020 research (grant agreement 771527- BROWSEC), by the Austrian Science Fund (FWF) through the projects PRO- FET (grant agreement P31621) and the project W1255-N23, by the Austrian Research Promotion Agency (FFG) through COMET K1 SBA and COMET K1 ABC, by the Vienna Business Agency through the project Vienna Cybersecurity and Privacy Research Center (VISP), by the Austrian Federal Ministry for Digital and Economic Affairs, the National Foundation for Research, Technology and Development and the Christian Doppler Research Association through the Christian Doppler Laboratory Blockchain Technologies for the Internet of Things (CDL-BOT).}
}

\author{\IEEEauthorblockN{Subhra Mazumdar}
\IEEEauthorblockA{\textit{TU Wien, Christian Doppler Laboratory} \\
\textit{Blockchain Technologies for the Internet of Things}\\
Vienna, Austria \\
subhra.mazumdar@tuwien.ac.at}
}

\maketitle 

\begin{abstract}
Hashed Timelock (HTLC)-based atomic swap protocols enable the exchange of coins between two or more parties without relying on a trusted entity. This protocol is like the \emph{American call option} without premium. It allows the finalization of a deal within a certain period. This puts the swap initiator at liberty to delay before deciding to proceed with the deal. If she finds the deal unprofitable, she just waits for the time-period of the contract to elapse. However, the counterparty is at a loss since his assets remain locked in the contract. The best he can do is to predict the initiator's behavior based on the asset's price fluctuation in the future. But it is difficult to predict as cryptocurrencies are quite volatile, and their price fluctuates abruptly. We perform a game theoretic analysis of HTLC-based atomic cross-chain swap to predict whether a swap will succeed or not. From the strategic behavior of the players, we infer that this model lacks fairness. We propose \emph{Quick Swap}, a two-party protocol based on hashlock and timelock that fosters faster settlement of the swap. The parties are required to lock griefing-premium along with the principal amount. If the party griefs, he ends up paying the griefing-premium. If a party finds a deal unfavorable, he has the provision to cancel the swap. We prove that \emph{Quick Swap} is more participant-friendly than HTLC-based atomic swap. Our work is the first to propose a protocol to ensure fairness of atomic-swap in a cyclic multi-party setting. 
\end{abstract}

\section{Introduction}
Centralized exchange enabled users to trade one cryptocurrency for the other. For example, \textit{Alice} wants to exchange $x_a$ coins for \textit{Bob}'s $y_b$ coins. The exchange can be done by involving a third party \emph{Carol}, where both \emph{Alice} can deposit $x_{a}+x$ coins and \emph{Bob} deposits $y_b$ coins with \emph{Carol} respectively. $x$ is the service fee charged by \emph{Carol} for offering the swap service. If \emph{Carol} is honest, she will handover $x_a$ coins to \emph{Bob}, $y_b$ coins to \emph{Alice} and keep the service charge. If she is malicious, she can just run away with \emph{Alice}'s and \emph{Bob}'s money. Thus , it become very important to have decentralized exchange of cryptocurrencies. With the introduction of Blockchain \cite{nakamoto2008bitcoin}, it is now possible to realize decentralized protocols for atomic swaps without relying on any trusted third party \cite{TN13,thomas2015protocol,H18,ZHLPGK19,zakhary2019atomic,TMM19,MBLNGK20,lys2021r,thyagarajan2021universal,narayanam2022generalized}. \emph{Alice} and \emph{Bob} can now safely lock their coins and have rules encoded in the smart contract. Exchange of assets leads to change of ownership, it either succeeds or fails in entirety. Bitcoin-based blockchains primarily leverage on Hashed Timelock Contracts or HTLC \cite{H18,borkowski2019dextt,dai2020research,narayanam2022generalized} for exchanging Bitcoins (BTC) with other cryptocurrencies like Litecoins (LTC), Ether (ETH), ERC Tokens.  

\begin{figure}[!ht]

    \centering
    \includegraphics[scale=0.34]{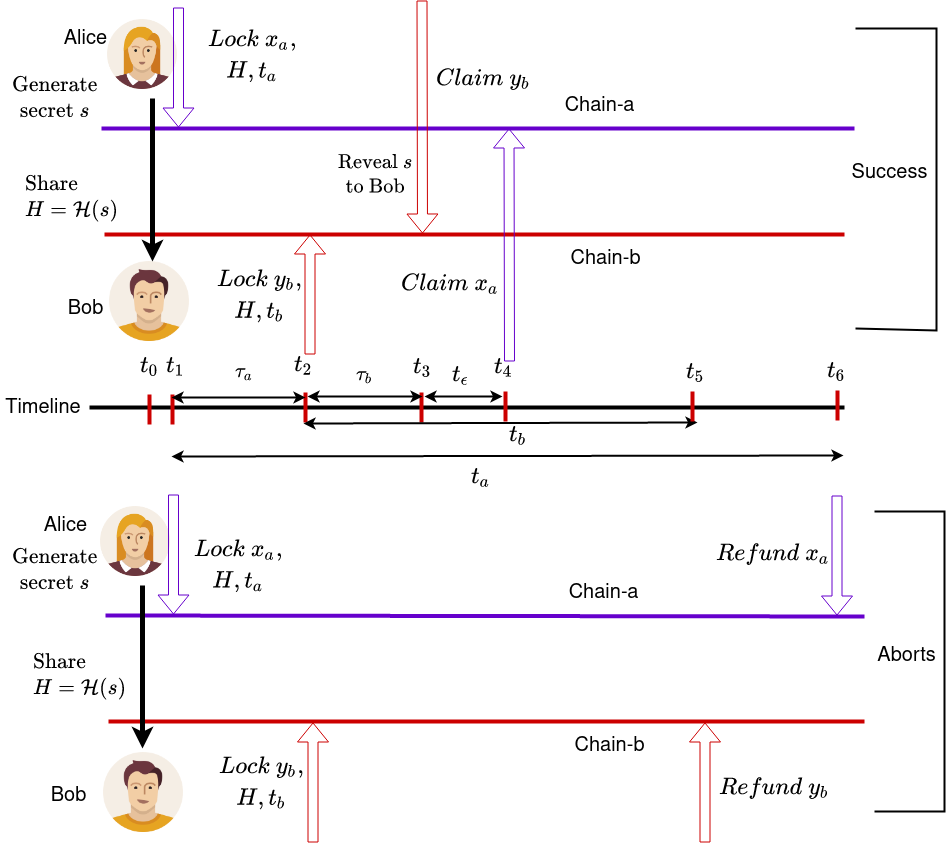}
    \caption{HTLC-based atomic swap. We assume that there is no waiting involved when parties lock their coins and are willing to exchange it as well.}
    \label{fig1}
\end{figure}

We explain a two-party HTLC-based atomic swap protocol with an example shown in Fig. \ref{fig1}. \emph{Alice} wants to exchange $x_a$ coins for $y_b$ coins of \emph{Bob}. Both of them have accounts in two different blockchains \texttt{Chain-a} and \texttt{Chain-b}. \emph{Alice} samples a secret $s$, generates a hash $H=\mathcal{H}(s)$ and shares it with \emph{Bob} at time $t_0$. The protocol comprises two phases: \emph{Lock} and \emph{Claim}. \emph{Lock phase} defines the time interval within which the parties have locked their assets in the contracts instantiated in the respective blockchains. \emph{Alice} locks $x_a$ coins in \texttt{Chain-a} at time $t_1$. The coin are locked in a contract where the spending conditions are as follows: \emph{if Bob provides the secret $s$ within $t_a$ units of time, then he claims $x_a$ coins else Alice initiates a refund after $t_a$ elapses}. Once the transaction is confirmed in \texttt{Chain-a} in the next $\tau_a$ units of time, \emph{Bob} locks $y_b$ coins in \texttt{Chain-b} at time $t_2$. He uses the same hash value $H$ in the contract deployed in \texttt{Chain-b} for locking his coins. The spending conditions are different: \emph{if Alice provides the secret $s$ within $t_b$ units of time, then she claims $y_b$ coins else Bob initiates a refund after $t_b$ elapses}, where $t_a>t_b$. It takes $\tau_b$ units for the lock transaction to be confirmed in \texttt{Chain-b}. \emph{Claim phase} signals the period within which the parties claim their assets. In the best case involving zero waiting, \emph{Alice} broadcasts the claim transaction at $t_3$, releases the preimage $s$ and claims $y_b$ coins from \emph{Bob}. The latter uses the preimage $s$ at broadcasts a transaction to claim $x_a$ coins at time $t_4$, where $t_4-t_3=t_{\epsilon}$ is the time taken by \emph{Bob} to observe \emph{Alice}'s transaction in \texttt{Chain-b}. Once the ownership of the assets changes successfully, an instance of the protocol succeeds. 

\subsection{Griefing Attack in Timelocked Contracts}

\begin{figure}[!ht]

    \centering
    \includegraphics[scale=0.34]{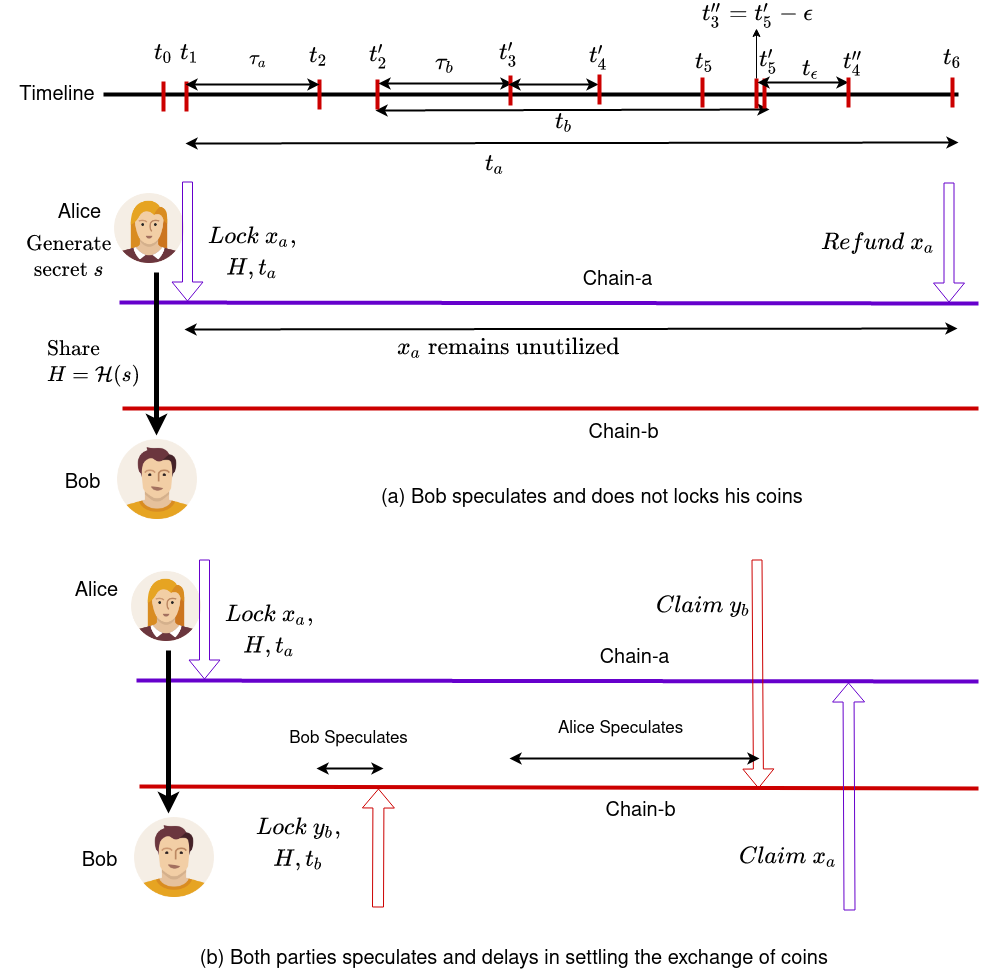}
    \caption{Coins remain unutilized in HTLC-based atomic swap due to (a) Bob speculates the deal and does not lock his coins and (b) Bob speculates, finds $t_2'$ favourable for locking his coins, delaying by $t_2'-t_2$, and Alice speculates, finds $t_3''=t_5'-\epsilon$ favourable for claiming the coins from Bob}
    \label{fig2delay}
\end{figure}

One of the main disadvantage of HTLC-based atomic swap is that parties are not enforced to settle the transaction. It has already been shown in \cite{han2019optionality}, \cite{zmn} that atomic swap is equivalent to \emph{American call option} without premium. In \emph{American call option}, buyer is allowed to exercise the contract no later than the strike time. We illustrate the situation in Fig. \ref{fig2delay}. After \emph{Alice} has locked $x_{a}$ coins at $t_1$, the time taken for the transaction to be confirmed in \texttt{Chain-a} is $\tau_a$. Ideally, \emph{Bob} should begin locking his coins at $t_2=t_1+\tau_a$. However, he may choose to delay speculate within duration $t_2$ to $t_6-(t_b+\epsilon+t_{\epsilon})$ and delay or he may choose to not lock his coins at all. If \emph{Bob} does not lock his coins, then \emph{Alice}'s coins get locked unnecessarily and she loses her coins in paying fees to miners for successful mining of refund transaction. This is termed as \emph{draining attack} \cite{eizinger2021open}. Let us assume that \emph{Bob} locks his coins at $t_2' \in [t_2,t_6-(t_b+\epsilon+t_{\epsilon})$. The time taken for the transaction to be confirmed in \texttt{Chain-b} is $\tau_b$. If \emph{Alice} chooses not to delay, then she will claim the coins at time $t_3'=t_2'+\tau_b$. However, \emph{Alice} may delay in claiming the coins or just abort. If she chooses to claim the coins at time $t_3''=t_5'-\epsilon$ where $t_5'=t_2'+\tau_b$ then \emph{Bob} gets to claim $x_a$ coins at $t_4''=t_3''+t_\epsilon$. So \emph{Bob} has to wait a duration of $t_4''-t_4'$ where $t_4'=t_3'+t_\epsilon$. Upon simplification, we observe that he waits $t_3''-t_3'=t_5'-\epsilon-t_3'$, which is the delay due to \emph{Alice}'s speculation. It may so happen that \emph{Alice} does not want to claim $y_b$ coins, she waits for $t_b$ units to elapse, and \emph{Bob} broadcasts refund transaction at time $t_5'=t_2'+t_b$. \emph{Alice} broadcasts her refund transaction after $t_a$ units elapse, i.e., at time $t_6=t_1+t_a$. If \emph{Alice} chooses not to respond, then it leads to a \emph{Griefing Attack} \cite{robinson2019htlcs}. Coins remain unutilized in either of the blockchains leading to substantial rise in opportunity cost. We define the attack formally.


\begin{definition}
\emph{(Griefing Attacks in Atomic Swap)} Given two parties \textbf{A} and \textbf{B}, such that \textbf{A} is required to forward an HTLC of $x_a$ coins to \textbf{B} for a certain timeperiod $t_a$, and in turn, \textbf{B} must forward an HTLC of $y_b$ coins to \textbf{A} for a timeperiod $t_b: t_a>t_b$, a griefing attack can happen in the following situations:
\begin{itemize}
    \item \emph{\textbf{A} locks $x_a$ coins and \textbf{B} doesn't lock $y_b$ coins}: It leads to $\textbf{A}'s$ $x_a$ coins being locked for $t_a$ units, the loss in terms of collateral cost being $\mathcal{O}(x_{a}t_a)$.
    \item \emph{$\textbf{A}$ locks $x_a$ coins, $\textbf{B}$ has locked $y_b$ coins, and $\textbf{A}$ aborts}: In such a situation, $\textbf{A}$ griefs $\textbf{B}$ at the cost of locking his coins for $t_a$ units of time. $\textbf{B}$'s coins remains locked for $t_b$ units, the loss in terms of collateral cost being $\mathcal{O}(y_{b}t_b)$.
\end{itemize}

\end{definition}

\subsection*{Motivation for Griefing in Atomic Swap}
A party may grief intentionally or decides to abort when the situation is not favourable for exchanging coins . We consider our parties to be either genuinely interested in exchanging coins or malicious. We define characteristic of each type:
\begin{itemize}
    \item \emph{Interested to Exchange}: A party who is willing to exchange coins but might end up griefing depending on whether she finds a favourable exchange rate.
    \item \emph{Malicious}: A party whose only motive is to mount Denial-of-Service (DoS) attack on the counterparty. Such a party will not take any actions after the counterparty has locked coins for exchange. Gain of malicious party is the lost opportunity cost of counterparty's locked coins.
\end{itemize}

Enforcing the parties not to back out from the deal is a major challenge. Additionally, we observe HTLC lacks flexibility as it does not provide an option to cancel the contract when the situation turns out to be unfavorable. \emph{Is it possible to propose an atomic swap that allows cancellation but at the same time penalizes malicious behavior?}

\subsection{ Contributions}
\begin{itemize}
    \item We model HTLC-based atomic cross-chain swap as a two-player sequential game and analyze the success rate of the protocol as a function of exchange rate and delay.
    \item We observe HTLC-based atomic swap is not participant-friendly by estimating the success rate of such a protocol.
    \item We propose \emph{Quick Swap}, a hashlock and timelock-based protocol compatible with Bitcoin script. Our protocol is more robust and efficient, allowing easy cancellation of trade and penalizing a party if it griefs. 
    \item \emph{Quick Swap} can also be generalized to multi-party cyclic atomic swap for countering griefing attacks.
\end{itemize}

\section{Game Theoretic Analysis of HTLC-based atomic swap}

\subsection{System Model \& Assumptions}
The atomic swap protocol comprises two phase: \emph{Lock Phase} - for the duration $t_0$ to just before $t_3$, and \emph{Claim Phase} - from time $t_3$ onward. It proceeds sequentially, with the assets being locked first and then the assets being claimed in the next phase. Given two parties Alice and Bob, their strategy space consists of the actions \emph{continue} and \emph{ stop}. In this paper, we will alternately use the term \emph{stop} and \emph{abort}, both denoting that a party chooses not to take any action. After \emph{Alice} locks $x_{a}$ coins in \texttt{Chain-a}, \emph{Bob} can choose to either \emph{continue}, i.e., lock $y_b$ coins in \texttt{Chain-b}, or abort from the process. There is no way by which \emph{Alice} can abort before timeout period $t_a$ elapses, if \emph{Bob} aborts. If \emph{Bob} locks his coins before that, \emph{Alice} can choose continue with the swap by claiming $y_b$ coins and releasing the preimage of $H$ before $t_b$ elapses. If she choose to \emph{stop}, she will wait for $t_b$ to elapse and \textit{Bob} initiates a refund after that. If either of the party is malicious, then the sole motive would be to keep the coins locked. So if \emph{Alice} is malicious, she will not initiate the \emph{Claim Phase}, and if \emph{Bob} is malicious, he will simply abort in the \emph{Lock Phase} without locking his coins.




\subsection{Basic Setup}
We denominate coins locked by \emph{Bob} as a function of value of \emph{Alice}'s coin at a given time $t$. Thus we express $y_b$ coins as $x(y_b,t)$, where the price is decided based on the exchange rate prevailing at time $t$. At time $t_0\approx t_1$, $x(y_b,t_1)$ is the price of \emph{Bob}'s coins decided to be exchanged for $x_a$ coins of \emph{Alice}. Price of \textit{Bob}'s coins at any time $t$ follow a geometric Brownian motion \cite{dipple2020using}. 
Coins locked by \textit{Bob} (or \textbf{B}), i.e., $y_b$, is denoted as $x(y_b,t)$, when denominated in terms of \textit{Alice}'s coins at time $t$. $x(y_b,t)$ follows a geometric Brownian motion:
\begin{equation}
\label{wiener}
    \ln{\frac{x(y_b,t+\tau)}{x(y_b,t)}}=\Big( \mu -\frac{\sigma^2}{2}\Big)\tau + \sigma (W_{t+\tau}-W_t)
\end{equation}
where $W$ follows a Wiener Process with drift $\mu$ and variance $\sigma^2$ \cite{Malliaris1990}. .\\\\
Given Eq.\ref{wiener}, the expected price of $x(y_b,t)$ at time $t+\lambda, (\mathcal{E}(x(y_b,t),\lambda)$, Probability density function of $x(y_b,t)$ at time $t+\lambda, (P(x,x(y_b,t), \lambda))$ , and Cumulative density function of $x(y_b,t)$ coins at time $t+\lambda, (C(x,x(y_b,t), \lambda))$ is expressed as follows:
\begin{equation}
    \begin{matrix}
        \mathcal{E}(x(y_b,t),\lambda)=\mathbb{E}[x(y_b,t+\lambda)|x(y_b,t)]=x(y_b,t)e^{\mu \lambda}\\\\
        P(x,x(y_b,t), \lambda)=\mathbb{P}[x(y_b,t+\lambda)=x|x(y_b,t)]\\
        =\frac{e^{-\Big(\frac{\ln{\frac{x}{x(y_b,t)}}-\Big(\mu -\frac{\sigma^2}{2}\Big)\tau}{2\tau\sigma^2}\Big)^2}}{\sqrt{2\pi\tau}\sigma x}\\\\
        C(x,x(y_b,t), \lambda)=\mathbb{C}[x(y_b,t+\lambda)\leq x|x(y_b,t)]\\
        =\frac{erfc{\Big(\frac{\ln{\frac{x}{x(y_b,t)}}-\Big(\mu -\frac{\sigma^2}{2}\Big)\tau}{\sqrt{2\tau}\sigma}\Big)}}{2}\\\\
    \end{matrix}
\end{equation}
where $erfc$ is the complementary error function. $erfc(x)$ is defined as:
\begin{equation}
    erfc(x) = \frac{2}{\Pi}\int_{x}^{\infty} e^{t^2} \,dt
\end{equation}
The expressions are taken from \cite{xu2021game}.

We assume both parties know each other's parameters. Transaction fees are assumed to be negligible compared to the amounts involved in the transactions. The rest of the notations are defined in Table \ref{tab:not}.
 \begin{table}[h]
\centering
  

  \scalebox{0.8}{
  \begin{tabular}{|c |c |} 
    \hline
 Notation &Description \\
 \hline
 $x_a$ &Coins possessed by \emph{Alice} or \textbf{A}\\
 $y_b$ & Coins of \emph{Bob} or \textbf{B} that \textbf{A} decides to buy for $x_a$ coins at time $t_0$\\
 $\tau_a$ &Time taken for a transaction to get confirmed in \texttt{Chain-a}\\
 $\tau_a$ &Time taken for a transaction to get confirmed in \texttt{Chain-b}\\
 $t_a$ &HTLC forwarded by \textbf{A} to \textbf{B} in \texttt{Chain-a}, locking $x_a$, expires\\
 $t_b$ &HTLC forwarded by \textbf{B} to \textbf{A} in \texttt{Chain-b}, locking $y_b$, expires\\
 $t_{\epsilon}$ &Propagation delay \\
 $\epsilon$ &Short time gap\\
 $\theta_1$ &Belief of \textbf{B} regarding type of \textbf{A} being \emph{interested to swap}\\
 $\theta_2$ &Belief of \textbf{A} regarding type of \textbf{B} being \emph{interested to swap} \\
 $sp_a$  &Success premium of \textbf{A}, if swap succeeds\\
 $sp_b$  &Success premium of \textbf{B}, if swap succeeds\\
$x(y_b,t)$ &Price of $y_b$ coins in terms of $x_a$ at time a given time $t$ \\
$r_a$  &Time discounting factor of \textbf{A}\\
$r_b$  &Time discounting factor of \textbf{B}\\
$f_a$ &Transaction fee in \texttt{Chain-a}\\
$f_b$ &Transaction fee in \texttt{Chain-b}\\
$\mu$ &Wiener process drift\\
$\sigma^2$ &Wiener process variance\\
$\mathcal{E}(x(y_b,t),\lambda)$ &Expected price of $x(y_b,t)$ at time $t+\lambda$, also expressed  \\
&as $\mathbb{E}[x(y_b,t+\lambda),x(y_b,t)]=x(y_b,t)e^{\mu \lambda}$\\
$P(x,x(y_b,t), \lambda)$ &Probability density function at time $t+\lambda$ \\
& given that price at time $t$ is $x(y_b,t)$\\
$C(x,x(y_b,t), \lambda)$ &Cumulative density function at time $t+\lambda$ \\
& given that price at time $t$ is $x(y_b,t)$\\

\hline

\end{tabular}
}
\vspace{0.2cm}
\caption{Notations used in the paper}
\label{tab:not}
\vspace{-0.2cm}
\end{table}

\subsection{Game Model}

 We model the interaction between two parties \emph{Alice}, denoted as \textbf{A}, and \emph{Bob}, denoted as \textbf{B}, as a sequential game $\Gamma_{swap}$. We assume that all the miners in the blockchains \texttt{Chain-a} and \texttt{Chain-b} are honest. In our model, \textbf{B} considers \textbf{A} to be \emph{interested in exchanging coins} with probability $\theta_1$ and \emph{malicious} with probability $1-\theta_1$. \textbf{A} considers \textbf{B} to be \emph{interested with probability} $\theta_2$ and \emph{malicious} with probability $1-\theta_2$. Given the belief \textbf{A} has regarding \textbf{B}'s type, if she has chosen to initiate the lock phase at time $t_1$ then \textbf{B} chooses either to lock $y_b$ coins or abort based on his belief of \textbf{A}'s type at time $t_2$. If \textbf{B} doesn't lock his coins before $t_1+t_{a}$ elapses then the swap stands canceled. If \textbf{B} locks his coins, \textbf{A} makes the next move at $t_3$, choosing either to claim $y_b$ coins or abort. Once \textbf{A} has chosen an action, \textbf{B} follows that. 


The extensive-form game $\Gamma_{swap}$, is defined for players \textbf{A} and \textbf{B}. The payoff function $u_{k, \theta}: S \times \mathbb{N} \rightarrow \mathbb{R}$ for any player $k \in \{\textbf{A,B}\}$ and $\theta \in \{\textrm{interested (int), malicious}\}$, where $S=\{continue, stop\}$, is denoted as $u_{k,\theta}(s,t)$ where $s \in S$ and $t\in \mathbb{N}$. $S$ denotes the set of actions for players \textbf{A} and \textbf{B}. $u_{k, \theta}(s,t)$ specifies the payoff the player $k$ of type $\theta$ would get at time $t\in \mathbb{N}$, if the players chose their action $s \in S$. The game begins with Nature \textbf{(N)} choosing the type of players \textbf{A} and \textbf{B}, where the probability of picking both players who are interested to exchange coins is $\theta_1\theta_2$ (probability of picking malicious \textbf{A} (\textbf{B}) is $1-\theta_1$ ($1-\theta_2$). Since selecting type of players are mutually independent, all possible combinations probability is the product of individual probability). In the next step, \emph{interested} \textbf{A} will selects her strategy of either \emph{continue} or \emph{stop}, based on the belief of \textbf{B}’s type, where as \emph{malicious} \textbf{A} will always choose to \emph{continue}. Next, \emph{interested} \textbf{B} chooses his strategy based on his belief of \textbf{A}'s type, however \emph{malicious} \textbf{B} will choose \emph{stop}. Finally, \textit{interested} \textbf{A} will choose to claim the coins if the exchange rates are in her favour else she aborts. \emph{Malicious} \textbf{A} will always abort.

\begin{figure*}[!ht]

    \centering
    \begin{subfigure}[b]{0.3\textwidth}
    \centering
    {\includegraphics[height=1.3in]{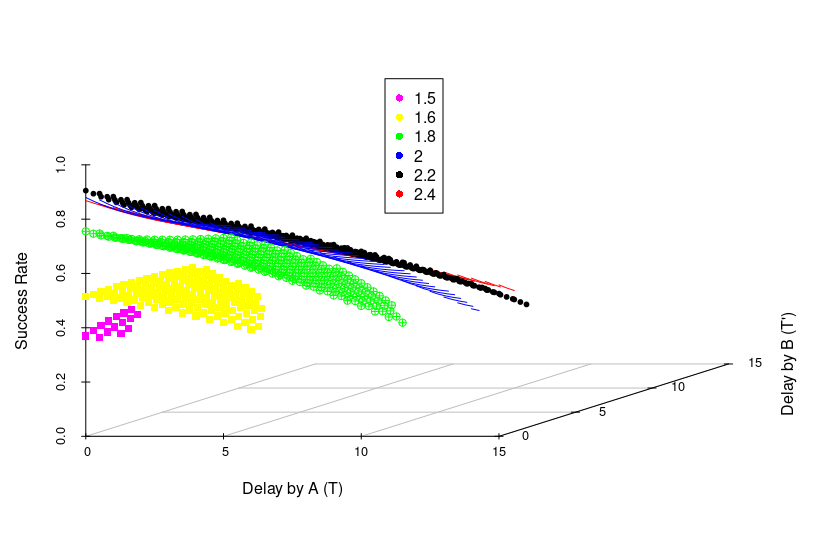}  }
    \label{march20}    
     \caption{$sp_a=sp_b=0.3,r_a=r_b=0.005,\sigma=0.1$} 
\end{subfigure}
    \begin{subfigure}[b]{0.3\textwidth}
    \centering
    {\includegraphics[height=1.3in]{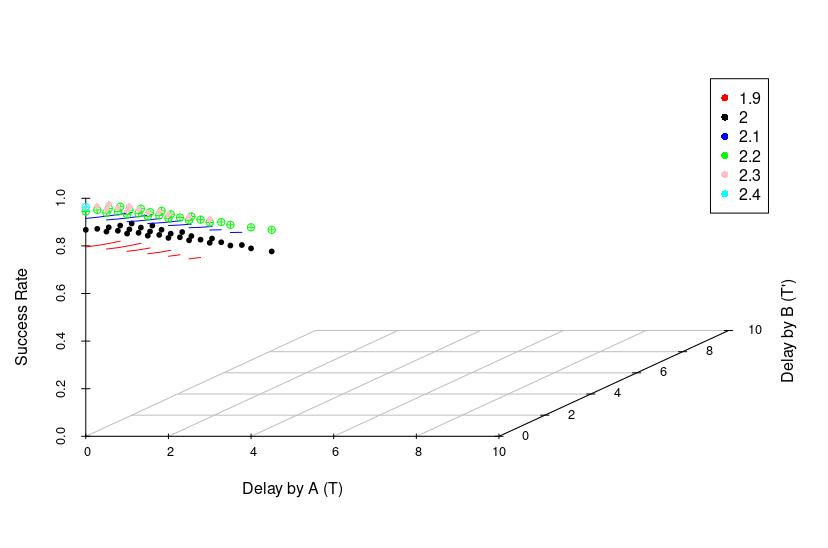} }   
    
    \label{april21}
     \caption{$sp_a=sp_b=0.3,r_a=r_b=0.01,\sigma=0.1$} 
\end{subfigure}
    \begin{subfigure}[b]{0.3\textwidth}
    \centering
{\includegraphics[height=1.3in]{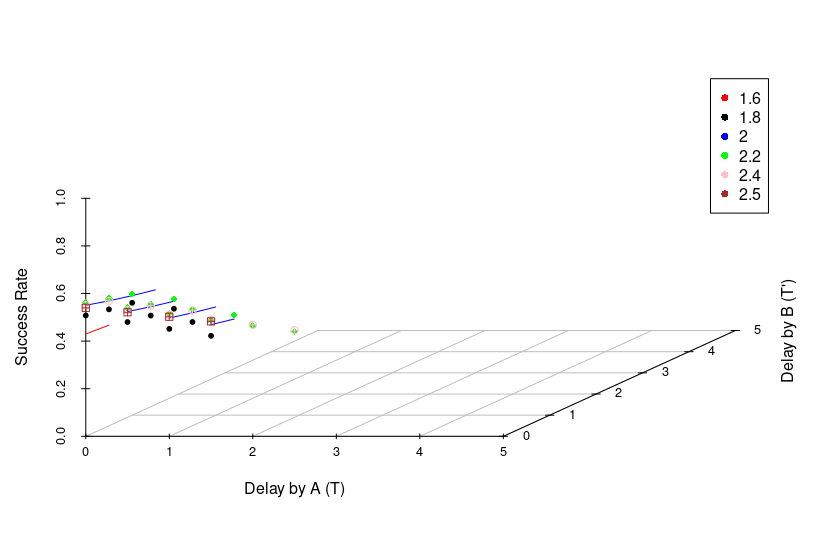}  }

     \caption{$sp_a=sp_b=0.3,r_a=r_b=0.005,\sigma=0.2$} 
\end{subfigure}




\caption{3-D plot of Swap Success Rate as a function of delay ($T,T'$) (x-y axis) and \textbf{A}'s coins (z-axis), with different parameter values, $\theta_1=\theta_2=0.5$}
\label{simulated}    
\end{figure*}

\subsubsection{Preference Structure}
To calculate the payoff of each party for type \emph{interested} \footnote{We do not calculate the payoff for type \emph{malicious} as we are interested in quantifying the success rate of HTLC-based atomic swap and this is possible only in presence of \emph{interested} parties.}, we use backward induction, starting from $t_3$. The claim phase starts at time $t_3$, when an \emph{interested} \textbf{A} decides whether to \emph{continue} and reveal preimage of $H$, or \emph{stop} for $t_b$ to elapse. If \emph{interested }\textbf{A} decides to claim $y_b$ coins, then an \emph{interested} \textbf{B} claims $y_b$ coins. We define the utility or payoff for each strategy. If \emph{interested} \textbf{A} chooses to \emph{continue} at $t_3$, then the time taken for the redeem transaction to get confirmed in \texttt{Chain-b} is $\tau_b$. We multiply payoff of \textbf{A} upon continuing with a factor $(1+sp_a)$, where $u_a$ is the success premium (or $u_{b}$ for \textbf{B}) to emphasize that rational parties gain higher utility by swapping their assets rather than abort. The utility of \emph{interested} \textbf{A} is expressed with their time discounted for duration $\tau_b$. Similarly, when \emph{interested }\textbf{B} claims coins in \texttt{Chain-a} at time $t_3+t_{\epsilon}$, time taken for the transaction to be confirmed is $\tau_a$, hence the utility is expressed with their time discounted for duration $t_{\epsilon}+\tau_a$.
\begin{equation}
    \begin{matrix}
    u_{A,int}(cont,t_3)=(1+sp_a) \mathcal{E}(x(y_b,t_{3}),\tau_b) e^{-r_a \tau_b}\\-f_b\\
    u_{B,int}(cont,t_3)= (1+sp_b) x_a e^{-r_b (\tau_a+t_{\epsilon})}-f_a \\

    \end{matrix}
\end{equation}

where $t_3=t_2+\tau_b+T$, and $T \in [0,t_b-\tau_b-\epsilon]$ being the delay by \emph{interested} \textbf{A} before she decides to claim the coins. If \emph{interested} \textbf{A} decides to \emph{stop}, then \emph{interested} \textbf{B} has to abort as well. The coins remain locked uselessly for the entire duration. The utility is expressed as: $u_{A,int}(stop,t_{3})= x_a e^{-r_a t_a}-f_a$ and $u_{B,int}(stop,t_{3})= \mathcal{E}(x(y_b,t_{3}),t_b) e^{-r_b t_b}-f_b$. An \emph{interested}  \textbf{A} will decide on \emph{continue} over \emph{stop} at $t_3$, if the following condition holds:
\begin{equation*}
\begin{matrix}
\label{s11}
(1+sp_a) \mathcal{E}(x(y_b,t_{3}),\tau_b) e^{-r_a \tau_b}-f_b \geq x_a e^{-r_a t_a}-f_a\\
or, x(y_b,t_3)e^{\mu\tau_b}\geq \frac{x_a e^{-r_a (t_a-\tau_b)}-(f_a-f_b)e^{r_a\tau_b}}{1+sp_a}\\
or, x(y_b,t_3)\geq \frac{\Big(x_a e^{-r_a (t_a-\tau_b)}-(f_a-f_b)e^{r_a\tau_b}\Big)e^{-\mu\tau_b}}{(1+sp_a)}\\
\end{matrix}
\end{equation*}
We derive $x(y_b,t_{3})^*$ in terms of $x_a$ for which the above inequality holds. If $x(y_b,t_3)\geq x(y_b,t_{3})^*$ then \textbf{A} claims the coins.

In the second round of \emph{lock phase}, \emph{interested} \textbf{B} has to decide whether he will \emph{continue} (i.e., lock $y_b$ coins) or \emph{stop} after \textbf{A} has locked $x_a$ coins. Since \emph{B} is unaware of \textbf{A}'s type, he calculates the expected payoff where with probability $1-\theta_1$ he has to \emph{stop} at time $t_2+\tau_b$ and with probability $\theta_1$, he can either \emph{continue}, if the price of coins rise to $x(y_b,t_3)^*$,  or \emph{stop} at time $t_2+\tau_b+T$, if the price drops. The payoff is expressed as time discounted, expected utility for duration $\tau_b+T$. An \emph{interested} \textbf{B} calculates what will be the probability that price of his coins will rise to $x(y_b,t_3)^*$ within time $\tau_b+T$. 

\begin{equation}
\label{Acalc}
\begin{matrix}

u_{B,int}(cont,t_{2})=\\ \theta_1 \Big( \frac{ \Big[ 1-C\Big(x(y_b,t_{3})^*,x(y_b,t_2),\tau_b+T\Big)\Big] u_{B,int}(cont,t_{3})  }{e^{r_b (\tau_b+T)}}  \\
+\\ \frac{\bigint_{0}^{x(y,t_{3})^*}  P\Big(p,x(y_b,t_2),\tau_b\Big) u_{B,int}(stop,t_{3}) \,dp}{e^{r_b (\tau_b)}} \Big)+\\ (1-\theta_1)\Big(\frac{u_{B,int}(stop,t_{3})}{e^{r_b\tau_b}}\Big)
\end{matrix}
\end{equation}
\emph{Interested} \textbf{A} knows what action was taken at $t_3$ and so there is no need to consider \textbf{B}'s type.
\begin{equation}
\label{Bcalc}
\begin{matrix}
 u_{A,int}(cont,t_{2})= \\ \frac{\bigint_{x(y_b,t_{3})^*}^{\infty}  P\Big(p,x(y_b,t_2),\tau_b+T\Big) u_{A,int}(cont,t_{3}) \,dp}{e^{r_a (\tau_b+T)}} \\+\frac{  C\Big(x(y_b,t_{3})^*,x(y_b,t_2),\tau_b\Big) u_{A,int}(stop,t_{3}) }{e^{r_a \tau_b}} \\
 


\end{matrix}
\end{equation}
If \textbf{B} \emph{stops} at $t_2$, then \textbf{A}'s coins remain locked for $t_a$ units of time. The utility is expressed as $u_{A,int}(stop,t_{2})= x_a e^{-r_at_a }-f_a$ and $u_{B,int}(stop,t_{2})= x(y_b,t_2)$.
\textbf{B}'s decision is dependent on how price of $y_b$ evolves until $t_{3}$. He will decide to \emph{continue} over \emph{stop}, if the following condition holds: $u_{B,int}(cont,t_2)\geq x(y_b,t_2)$. We derive the range in which $x(y_b,t_2)$ must lie for the above inequality to hold. \textbf{B} will continue if $x_1<x(y_b,t_2)\leq x_2$. If $x(y_b,t_2)>x_2$ or $x(y_b,t_2)\leq x_1$ then \textbf{B} will stop.

In the first round of the lock phase,  \emph{interested} \textbf{A} takes a decision on whether she wants to lock coins or abort at time $t_1$ based on her belief regarding \textbf{B}'s type. \textbf{B} can choose to take action at $t_2=t_1+\tau_a+T'$ where $T' \in [0,t_a-( t_b -\epsilon+t_{\epsilon})-\tau_a]$. Payoff of \textbf{A} can be expressed with their time-discounted, expected utility for duration $t_{2}-t_1$.
\begin{equation}
\begin{matrix}
u_{A,int}(cont,t_{1})= \\ \theta_2\Big(\frac{\bigint_{x_1}^{x_2}  P\Big(p,x(y_b,t_1),\tau_a+T'\Big) u_{A,int}(cont,t_{2}) \,dp}{e^{r_a(\tau_a+T')}} +\\  
\frac{ \Big[1- C\Big(x_2,x(y_b,t_1),\tau_a\Big)+C\Big(x_1,x(y_b,t_1),\tau_a\Big) \Big] u_{A,int}(stop,t_{2}) }{e^{r_a\tau_a}}\Big)+\\(1-\theta_2)\Big(\frac{u_{A,int}(stop,t_2)}{e^{r_a\tau_a}}\Big)
\end{matrix}
\end{equation}
\emph{Interested} \textbf{B} knows what action was taken at $t_2$ and so there is no need to consider \textbf{A}'s type.

\begin{equation}
\begin{matrix}
u_{B,int}(cont,t_{1})= \\ \frac{\bigint_{x_1}^{x_2}  P\Big(p,x(y_b,t_1),\tau_a+T'\Big) u_{B,int}(cont,t_{2}) \,dp}{e^{r_b(\tau_a+T')}}+\\ \frac{\bigint_{0}^{x_1}  P\Big(x,x(y_b,t_1),\tau_a\Big) u_{B,int}(stop,t_{2}) \,dp}{e^{r_b\tau_a}}

\end{matrix}
\end{equation}
If \emph{interested} \textbf{A} stops at $t_1$, then no one locks coins and we express the utility as $u_{A,int}(stop,t_{1})= x_a$ and $u_{B,int}(stop,t_{1})= x(y_b,t_1)$. \textbf{A}'s decision is based on how price of $y_b$ evolves until $t_{2}$. \textbf{A} will decide on \emph{continue} over \emph{stop}, if the following condition holds: $u_{A,int}(cont,t_{1})\geq x_a$. We derive the range $(x^*,x^{*'})$, in which $x_a$ must lie for the swap to start.

\begin{proposition}
HTLC-based atomic swap is not participant-friendly.
\end{proposition}
\emph{Proof}: \textbf{A}'s willingness to participate in the atomic swap is decided by the expected success of the protocol for given set of parameters. The success rate \emph{(SR)} of a swap is the probability that the swap succeeds after it has been initiated,
i.e. after \textbf{A} has locked coins at $t_1$ \cite{xu2021game}. For a given pair of $\theta_1$ and $\theta_2$. \emph{SR} is defined as function of $x_a$ (\textbf{A}'s coins or tokens), delay by \textbf{A} (T) at final step while claiming coins, and delay by \textbf{B} (T') at second step while locking $y_b$ coins. It is expressed as:
\begin{equation}
\begin{matrix}
    SR(x_a,T',T)= \bigints_{x_1[x_a]}^{x_2[x_a]} P_A(p,T')P_B(p,T)\, dp
    \end{matrix}
\end{equation}
where $ P_A(p,T')=\theta_2 P(p,x(y_b,t_1),\tau_a+T')$ and $P_B(p,T)=\theta_1\Big(1-C(x(y_b,t_3)[x_a],p,\tau_b+T)\Big)$. 

We plot the success rate of the protocol in Fig. \ref{simulated} (a-c), the parameters used are $t_\epsilon\approx \epsilon = 1 \ hr$ and $\tau_a=\tau_b=3 \ hrs$.  As per standard practice, $t_a = 48 \ hrs$ and $t_b=24 \ hrs$ \cite{han2019optionality}. We select $x_a \in [1,3]$ given $x(y_b,t_1)=2$. The parameters $T$ is varied between $[0,20$] and $T'$ is varied between $[0,21]$. The fee $f_a$ and $f_b$ is negligible, so we consider them to be 0. The success premiums $sp_a=sp_b=0.3$, time-discounting factor $r_a=r_b$ is chosen from $\{0.005,0.01\}$, $\sigma$ is varied between 0.1 and 0.2, and $\mu$ is selected from $\{-0.002,0.002\}$. We observe that the success rate is $ \geq 0.9$ in Fig. \ref{simulated} (a-b), and around 0.6 for Fig. \ref{simulated} (c), when $T=T'=0$. When $T$ or $T'$ increases, the success rate drops and beyond certain range, it becomes $NA$ (not applicable) as $u_{A_{continue},t_{1}}< x_a$.
Success rate is function of $T$ and $T'$ that cannot be determined before the swap proceeds to the second round or to the third round. In the worst case, if $T\approx t_b$ and/or $T'\approx t_a-t_b$, then the success rate drops drastically as the payoff upon continuing is too low. In such situation, a party is better off if he does not participate in the swap rather than keep his coins locked for one full day.

\section{Quick Swap: A protocol based on Ideal Swap Model}
The flaw in the HTLC-based atomic swap is that either of the parties can speculate and delay in settling the transaction without losing anything. Our objective is to force the parties to settle the transaction faster, and penalize for delayed action. We provide a high-level overview of our proposed protocol where a party will lock the principal amount for swap provided he or she gets a guarantee of compensation upon suffering from a griefing attack. With respect to the previous example, \emph{Alice} has to lock $x_a$ coins for a period of $t_a$ units. If \emph{Bob} griefs, \emph{Alice}'s collateral locked will be $\mathcal{O}(x_at_a)$. She calculates the collateral cost of locking $x_{a}$ coins for $t_a$, let it be defined as $c(x_{a}t_a)$. Similarly, \emph{Bob} calculates the collateral cost of locking $y_{b}$ coins for $t_b$, let it be $c(y_{b}t_b)$. The steps for locking the coins proceeds as follows:
\begin{itemize}
    \item[(i)] \emph{Bob} first locks $c(x_{a} t_a)$ coins in \texttt{Chain-b} for $D+\Delta$ units of time where $\tau_a+2\tau_b< D+\Delta< t_b$. We will explain later why an additional $\Delta$ unit is chosen.
    \item[(ii)] After \emph{Alice} gets a confirmation of the griefing-premium locked by \emph{Bob}, she locks $x_{a}$ for $t_a$ units and griefing-premium $c(x_at_a)+c(y_{b} t_b)$ for $D>\tau_a+\tau_b$ units of time, both in \texttt{Chain-a} .
    \item[(iii)] After \emph{Bob} gets a confirmation of the griefing-premium locked by \emph{Alice}, he locks $y_{b}$ for $t_b$ units of time in \texttt{Chain-b}.
    
\end{itemize}

If \emph{Bob} doesn't want to proceed after step [ii] and cancels the swap, then \emph{Alice} can unlock $x_a$ coins from \texttt{Chain-a}. If \emph{Bob} griefs, then \emph{Alice} gets the compensation $c(x_at_a)$ after $D+\Delta$ elapses instead of being griefed for $t_a$. The principle we follow here is \emph{``Coins you have now is better than coins you have later''}. \emph{Alice} can use the compensation from $t_b-(D+\Delta)$. Had we set the timelock for locking $c(x_at_a)$ to $t_a$, \emph{Bob} can still grief by canceling the contract at time $t_b-\epsilon$. We will discuss later how should we choose $D$ to ensure a faster compensation.

If \emph{Alice} initiates the swap, she claims $y_b$ coins and withdraws the compensation from \texttt{Chain-a}. \emph{Bob} gets to claim $x_a$ coins and he withdraws the compensation from \texttt{Chain-b}. If \emph{Alice} cancels the swap, then \emph{Bob} unlocks $y_b$ coins from \texttt{Chain-b}. If \emph{Alice} delays beyond $D$ unit of time, then \emph{Bob} gets compensation for the loss. Since $D+\Delta<\tau_b$, \emph{Alice} can delay beyond this point as well. In that case, she gets a compensation of $c(x_at_a)$ and her net gain is $g_a=c(x_at_a)-c(y_bt_b)$ and \emph{Bob}'s net loss is $-g_a$. Thus to prevent \emph{Bob} from incurring a loss, \emph{Alice} is forced to pay a compensation of $c(x_at_a)+c(y_bt_b)$. Even if she does not respond, \emph{Bob} is entitled to a compensation of $c(y_bt_b)$ after refunding $c(x_at_a)$ coins to \emph{Alice}. 



\subsection{Formal Description of Quick Swap}
\subsubsection{System Model and Assumption}
The system model and assumptions are same as HTLC-based atomic swap. Since $t_a \approx 2 t_b$, for ease of analysis, we consider $c(x_at_a)\geq 2c(y_bt_b)$. For a fixed rate of griefing-premium, we consider $Q=c(x_at_a)$ and $\frac{Q}{2}=c(y_bt_b)$. \emph{Alice} locks a griefing-premium of 1.5Q and \emph{Bob} locks griefing-premium Q coins. We denote \emph{Alice} as \textbf{A} and \emph{Bob} as \textbf{B} while describing the protocol.
%
%
\subsubsection{Detailed Description}
\label{def}
\begin{figure}[!ht]
    \centering
 
    \includegraphics[scale=0.34]{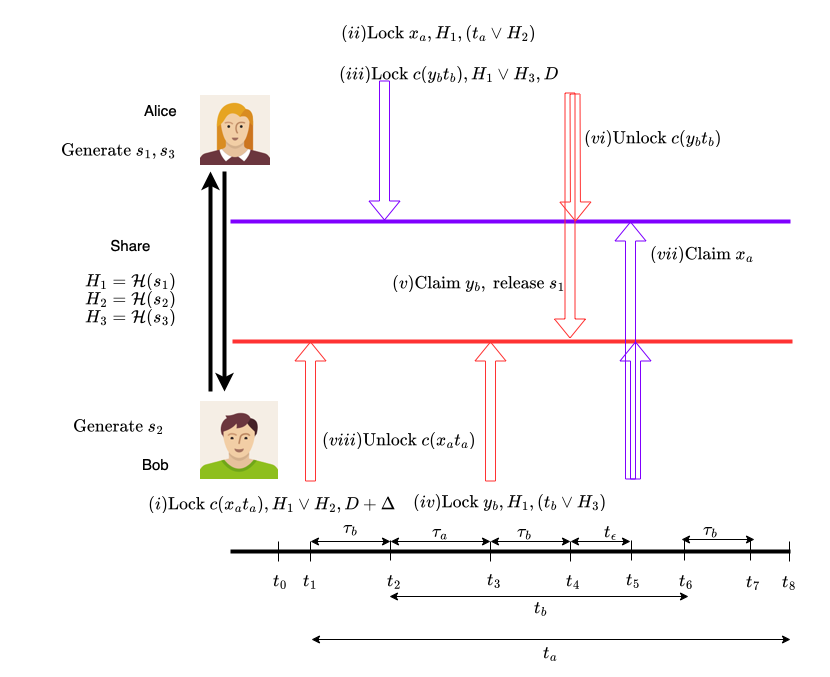}
    \caption{An instance of successful Quick Swap between Alice and Bob}
    \label{figswap}
\end{figure}

The protocol has the following phases: (A) Preprocessing Phase, (B) Lock Phase and (C) Claim Phase. An instance of successful execution of \emph{Quick Swap} is shown in Fig. \ref{figswap}. 


(A) \textbf{Preprocessing Phase}: The steps are defined as follow:\\
 \textit{Sampling Cancellation Hash and Payment Hash}:
\begin{itemize}
\item[(i)] \textbf{A}'s pair of secret and public key is $(sk_a,pk_a)$ and \textbf{B}'s pair of secret and public key is $(sk_b,pk_b)$. \textbf{A} uses $sk_a$ and \textbf{B} uses $sk_b$ for signing transactions. $pk_a$ and $pk_b$ are used for verifying each such signed transaction.
\item[(ii)] \textbf{A} samples random values $s_1$ and $s_3$, creates payment hash $H_1=\mathcal{H}(s_1)$ and cancellation hash $H_3=\mathcal{H}(s_3)$. She shares $H_1$ and $H_3$ with \textbf{B}.
\item[(ii)] \textbf{B} samples a random value $s_2$ and creates cancellation hash $H_2=\mathcal{H}(s_2)$ and shares it with \textbf{A}.
\end{itemize}

(B) \textbf{Locking Phase}:

\begin{itemize}

\item[(i)] At time $t_1$, \textbf{B} creates and signs transaction \texttt{griefing\_premium\_lock\_B} using funding address $addr_{funding,penalty,B}$ that locks $Q$ coins address $addr_{lock,penalty,B}$ and publishes \texttt{griefing\_premium\_lock\_B} in \texttt{Chain-b}. \textbf{B}  encodes the condition whereby $Q$ coins can be claimed either by revealing the preimage of $H_1$ or $H_2$, or it can be spend by \textbf{A} after $D+\Delta$ units of time. 
\item[(ii)] \textbf{A} checks whether \texttt{griefing\_premium\_B} is confirmed within $\tau_b$ units. Once that is confirmed, at time $t_2=t_1+\tau_b$, \textbf{A} creates and signs a transaction \texttt{griefing\_premium\_lock\_A} using funding address $addr_{funding,penalty,A}$ that locks $1.5Q$ coins into address $addr_{lock,penalty,A}$. The condition for spending the coins are as follows:\\
(a) Either provide the preimage of $H_1$ or $H_3$;\\
(b) If no preimage is provided, \textbf{B} can spend the coins locked after $D$ units of time.\\
 Simultaneously, \textbf{A} creates and signs another transaction \texttt{principal\_lock\_A} using funding address $addr_{funding,principal,A}$ that locks $x_a$ coins into address $addr_{lock,principal,A}$. The coins can be spend either by revealing the preimage of $H_1$, or, \textbf{A} can refund the coins either after $t_a$ unit elapses or by revealing preimage of $H_2$, whichever occurs first. She publishes both \texttt{principal\_lock\_A} and \texttt{griefing\_premium\_lock\_A} in \texttt{Chain-a}. 
\item[(iii)] \textbf{B} checks whether the transactions broadcasted by \textbf{A} gets confirmed in another $\tau_a$ units. At time $t_3=t_2+\tau_a$, He creates and signs another transaction \texttt{principal\_lock\_B} using funding address $addr_{funding,principal,B}$ that locks $y_b$ coins into address $addr_{lock,principal,B}$. The coins can be spend either by revealing the preimage of $H_1$, or, \textbf{B} can refund the coins either after $t_b$ unit elapses or by revealing preimage of $H_3$, whichever occurs first. He then proceeds to publish \texttt{principal\_lock\_B} in \texttt{Chain-b}. 
\end{itemize}
(C) \textbf{Claim Phase}: Once \textbf{A} observes that \textbf{B} has locked $y_b$ coins in \texttt{Chain-b}, he initiates the claim phase at time $t_4=t_3+\tau_b$, where $tau_b$ is the time taken for \texttt{principal\_lock\_B} to be confirmed. \\
\emph{Redeem}: If \textbf{A} wishes to redeem the coins,\\
a. At time $t_4=t_3+\tau_b<D$:
\begin{itemize}
    \item[(i)] \textbf{A} releases the preimage $s_1$ for payment hash $H_1$ and claims the output of transaction \texttt{principal\_lock\_B} with her signature in \texttt{Chain-b}. This allows her to claim $y_b$ coins.
\item[(ii)]  \textbf{A} uses $s_1$ to refund the griefing-premium of $1.5Q$ locked in \texttt{Chain-a}.  
\end{itemize}

b. At time $t_5=t_4+t_{\epsilon}$: 
\begin{itemize}
    \item[(i)] \textbf{B} uses the preimage $s_1$ and unlocks the output of the transaction \texttt{principal\_lock\_A} with his signature in \texttt{Chain-a}, claiming $x_a$ coins.
\item[(ii)]  \textbf{B} uses $s_1$ to refund $Q$ locked in \texttt{Chain-b}.  
\end{itemize}

\textit{Refund}: If \textbf{A} wishes to cancel the swap, \\
a. At time $t_4=t_3+\tau_b<D$:
\begin{itemize}
    \item[(i)] \textbf{A} releases the preimage $s_3$ for cancellation hash $H_2$ with her signature in \texttt{Chain-a}, unlocking $1.5Q$ coins.
\item[(ii)]  \textbf{B} uses $s_3$ to refund $y_b$ coins from \texttt{Chain-b} at time $t_4+t_{\epsilon}$. This results in cancelation of swap before $t_b$ elapses.  
\end{itemize}

b. At time $t_5=t_4+\tau_b+t_{\epsilon}$:  
\begin{itemize}
    \item[(i)] \textbf{B} releases the preimage $s_2$, unlocks $Q$ coins with her signature from \texttt{Chain-b}.
\item[(ii)]  \textbf{A} uses $s_2$ for refunding $x_a$ coins from \texttt{Chain-a} at time $t_4+\tau_b+2t_{\epsilon}$.  
\end{itemize}

\subsection{Proof of Correctness, Safety and Liveness}
It is necessary to argue the state of a proposed protocol in presence of both compliant and malicious parties. Parties either may choose to follow the protocol or they may deviate. We prove that \emph{Quick Swap} satisfies both safety and liveness. By safety, we mean that \emph{``compliant parties should end up “no worse off,” even when other parties deviate arbitrarily from the protocol''} \cite{herlihy2022cross}. Simultaneously, the liveness property states that none of the parties end up keeping their coins locked forever. Before arguing for liveness and safety, we prove the correctness of the protocol in presence of compliant parties - either all agreed coin exchange take place, and no exchange take place. 
\begin{property}
(\emph{Correctness}) If all parties are compliant, then the swap either succeeds with coins being exchanged, or the swap gets canceled with the coins being refunded.
\end{property}
\emph{Proof}: \textbf{A} and \textbf{B} exchange the hashes $H_1,H_2$ and $H_3$ before the start of the protocol. At $t_1$, \textbf{B} locks $Q$ coins in \texttt{Chain-b} that can be unlocked contingent to either providing preimage of $H_1$ or $H_2$. The coins are locked for $D+\Delta$ units, after which \textbf{A} can claim $Q$ coins and begins the next phase of locking at $t_2=t_1+\tau_b$. She locks $x_a$ coins in \texttt{Chain-a} that can be claimed by \textbf{B} contingent to providing the preimage of $H_1$, else \textbf{A} refunds after time $t_a$ or using preimage of $H_2$. She also locks $1.5Q$ coins at $t_2$ in \texttt{Chain-a}. The amount can be unlocked contingent to either providing preimage of $H_1$ or $H_3$. The coins are locked for $D$ units, after which \textbf{B} can claim $1.5Q$ coins. The last locking phase starts at $t_3=t_2+\tau_a$, when \textbf{B} locks $y_b$ coins in \texttt{Chain-b}. The latter can be claimed by \textbf{A} contingent to providing the preimage of $H_1$, else \textbf{B} refunds after time $t_b$ or using preimage of $H_3$. The correctness of \emph{Claim Phase} follows from the description of \emph{Redeem} and \emph{Refund} defined in Section \ref{def}. 
  
\begin{property}
(\emph{Safety Property}) The safety property states that
compliant parties should be as better off as they had been before the protocol execution, even when
other parties deviate arbitrarily from the protocol.
\end{property}
\emph{Proof}: After $\textbf{B}$ has locked $Q$ coins \texttt{Chain-b}, if $\textbf{A}$ does not lock any coins \texttt{Chain-a}, $B$ unlocks it after a certain timeperiod $\delta<D+\Delta$.
If $\textbf{B}$ locks $Q$ coins, $\textbf{A}$ locks $x_a+1.5Q$ coins, but $\textbf{B}$ aborts without locking $y_b$ coins, then $\textbf{A}$ unlocks $1.5Q$ coins revealing preimage $s_3$. She gets a compensation $Q$ coins after $D+\Delta$ that covers up for the lost opportunity cost of keeping $x_a$ coins locked for $t_a$ units.
If all the parties have locked their coins but $\textbf{A}$ delays beyond $D$ or griefs, $\textbf{B}$ gets compensation of $1.5Q$. He may lose $Q$ coins (if $\textbf{A}$ cancels before elapse of $D+\Delta$ but after timeout $D$, then $\textbf{B}$ is entitled to the full compensation $1.5Q$) but we ensure that $1.5Q-Q=0.5Q$ coins are enough to compensate for locked collateral $\mathcal{O}(y_bt_b)$. 

\begin{property}
\emph{(Liveness)} Coins No asset belonging to a compliant party do not remain locked forever.
\end{property}
\emph{Proof}: If \textbf{A} doesn't take any action by $t_2=t_1+\tau_b$, \textbf{B} unlocks $Q$ coins after $\tau_b> \delta>0$ units. 
If \textbf{B} does not lock coins at $t_3=t_2+\tau_a$, then by time $t_3+\tau_b$, \textbf{A} refunds the griefing-premium $1.5Q$ by revealing preimage of $H_3$ in her own interest. \textbf{B} observes that \textbf{A} has canceled the swap by withdrawing the griefing-premium. If he is rational, then he will releases preimage $s_2$ for $H_2$, unlock $Q$ coins and allow \textbf{A} to unlock $x_a$ coins before $D$ elapses. If \textbf{B} is malicious, then he will end up losing $Q$ coins after $D+\Delta$ units, but \textbf{A} will be able to unlock $x_a$ coins after $t_a$ elapses. 
If at time $t_4=t_3+\tau_b$, \textbf{A} aborts then she loses compensation of $1.5Q$ to \textbf{B}. The latter can unlock $y_b$ coins after $t_b$ has elapsed but loses his compesation of $Q$ coins. 
 

%

\subsection{Game-Theoretic Analysis}
 We model the interaction between the two entities \textbf{A} and \textbf{B} as a sequential game $\Gamma_{quick \ swap}$. \textbf{B} initiates the lock phase by locking griefing-premium $Q$ for duration $D+\tau_a+\tau_b+\Delta$ units in \texttt{Chain-b} at time $t_1$, followed by \textbf{A} choosing either to lock $x_a$ coins for duration $t_a$ units or abort at $t_2=t_1+\tau_b$. If \textbf{A} has not responded then \textbf{B} either cancels the swap at time $t_2+\tau_a$ by unlocking $Q$ or he stops. If \textbf{A} wishes to continue, she locks the principal amount and the griefing-premium $1.5Q$ for $D$ units in \texttt{Chain-a} at time $t_2$. At time $t_3=t_2+\tau_a$, if \textbf{B} observes that \textbf{A} has locked the principal amount as well as griefing-premium, then \textbf{B} either locks $y_b$ coins for $t_b$ units in \texttt{Chain-b} or stops. If he does not lock coins, then at time $t_3+\tau_b$, \textbf{A} cancels the swap by unlocking $1.5Q$ coins from \texttt{Chain-a} or stops. If \textbf{B} has not responded, he loses the griefing-premium at time $t_1+t_a$. Else if he chooses to cancel, then \textbf{A} will be able to withdraw the principal amount as well. If \textbf{B} has chosen to continue, \textbf{A} decides at time $t_4=t_3+\tau_b$, whether to continue or cancel the swap. 
 \subsubsection{Game Model \& Preference Structure}
The extensive-form game $\Gamma_{quick \ swap}$ is similar to $\Gamma_{swap}$, except that \textbf{A} and \textbf{B}'s strategy space has the option to \emph{cancel}, apart from \emph{continue} and \emph{stop}. The analysis is done by applying backward induction on $\Gamma_{quick \ swap}$.

If \textbf{A} delays instead of making a move at $t_3+\tau_b$, then the opportunity cost of coins locked as griefing-premium will rise. Canceling the swap at $t_3+\tau_b$ will lead to \textbf{A}'s utility as $\frac{x_a}{e^{r_a(t_{\epsilon}+2\tau_a)}}+\frac{1.5Q}{e^{r_a \tau_a }}$. If \textbf{A} chooses to delay and cancel at time say $t_3+\tau_b+t$ for $t>0$, then the utility drops further, i.e, $\frac{x_a}{e^{r_a(t_{\epsilon}+2\tau_a+t)}}+\frac{1.5Q}{e^{r_a (\tau_a+t) }}$. In the previous HTLC-based atomic swap, if \textbf{A} finds that the utility on continuing is less than the utility of the swap till the lock time of HTLC expires, she speculated till the situation turns in her favor. However, the situation is different now as \textbf{A} is allowed to abort the swap much earlier without waiting for the lock time to elapse. A rational \textbf{A} will choose not to delay anticipating that the situation may turn worse later. At time $t_4$, if \textbf{A} continues and \textbf{B} follows:
\begin{equation}
    \begin{matrix}
u_{A,int}(cont,t_{4})=(1+sp_a) \frac{\mathcal{E}(x(y_b,t_4),\tau_b)}{e^{r_a\tau_b}}+ \frac{1.5Q}{e^{r_a \tau_a }}-f_a-f_b\\
     u_{B,int}(cont,t_{4})= (1+sp_b)  \frac{x_a}{e^{r_b (\tau_a+t_{\epsilon})}}+ \frac{Q}{e^{r_b(t_{\epsilon}+\tau_b)}}\\-f_a-f_b
    \end{matrix}
\end{equation}
At time $t_4$, if \textbf{A} cancels then \textbf{B} cancels the deal as well. The payoffs are $u_{A,int}(cancel,t_{4})= \frac{x_a}{e^{r_a(2t_{\epsilon}+\tau_b+\tau_a)}}+\frac{1.5Q}{e^{r_a \tau_a }}-2f_a$ and
     $u_{B,int}(cancel,t_{4})=  \frac{\mathcal{E}(x(y_b,t_4),t_{\epsilon}+\tau_b)}{e^{r_b (\tau_b+t_{\epsilon})}}+\frac{Q}{e^{r_b(t_{\epsilon}+2\tau_b)}}-2f_b$. 
\textbf{A} will continue at $t_4$ over canceling the swap, if the following condition holds:
\begin{equation}
\begin{matrix}
(1+sp_a) \frac{\mathcal{E}(x(y_b,t_4),\tau_b)}{e^{r_a\tau_b}}+ \frac{1.5Q}{e^{r_a \tau_a }}-f_a-f_b> \frac{x_a}{e^{r_a(2t_{\epsilon}+\tau_b+\tau_a)}}\\+\frac{1.5Q}{e^{r_a \tau_a }}-2f_a\\\\
or, \ x(y_b,t_4)> \frac{\frac{x_a}{e^{r_a(2t_{\epsilon}+\tau_b+\tau_a)}}-f_a+f_b}{(1+sp_a)e^{(\mu-r_a)\tau_b}}
\end{matrix}
\end{equation}
   We derive $x(y_b,t_{4})^*$ in terms of $x_a$ for which the above inequality holds. If $x(y_b,t_4)\geq x(y_b,t_{4})^*$ then \textbf{A} claims the coins.

     

At time $t_3$, \textbf{B} decides to continue, then with probability $1-\theta_1$, \textbf{A} is malicious and will delay till $D-\epsilon$ ($D-\epsilon \rightarrow D$). The utility is expressed as follows:
$u_{B,int}(cont,t_{3})=\theta_1\Big(\frac{ \Big[ 1-C(x(y,t_{4})^*,x(y,t_3),\tau_b)\Big] u_{B,int}(cont,t_{4}) \,dp}{e^{r_b \tau_b}}+ \\ \frac{\bigint_{0}^{x(y,t_{4})^*}  P(p,x(y,t_3),\tau_b) u_{B,int}(cancel,t_{4}) \,dp}{e^{r_b \tau_b}}\Big)+(1-\theta_1)\Big( \frac{\mathcal{E}(x(y_b,t_4),t_{\epsilon}+D)}{e^{r_b (D-\epsilon+t_{\epsilon})}}+\frac{Q}{e^{r_b(t_{\epsilon}+D-\epsilon+\tau_b)}}-2f_b \Big)$ and $u_{A,int}(cont,t_{3})= \\ \frac{\bigint_{x(y,t_{4})^*}^{\infty}  P(p,x(y,t_3),\tau_b) u_{A,int}(cont,t_{4}) \,dp}{e^{r_a \tau_b}}+  \frac{\bigint_{0}^{x(y,t_{4})^*}  P(p,x(y,t_3),\tau_b) u_{A,int}(cancel,t_{4}) \,dp}{e^{r_a \tau_b}}$
 

At time $t_3$, \textbf{B} chooses stop, then the utility for \textbf{A} and \textbf{B} is defined as: 
$u_{A,int}(stop,t_{3})=\frac{x_a}{e^{r_a(t_a+\tau_a)}}+\frac{1.5Q}{e^{r_a (t_b+\tau_a) }}-2f_a+ \frac{Q}{e^{r_b\tau_b}}-f_b$ and $u_{B,int}(stop,t_{3})=x(y_b,t_3)$.

If at time $t_3$, \textbf{B} chooses cancel then he unlocks the premium $Q$ locked in \texttt{Chain-b} by releasing preimage of $H_2$. \textbf{A} observes that swap is canceled, so she unlocks $x_a$ coins and griefing-premium $1.5Q$ from \texttt{Chain-a}. The utility is defined as $u_{B,int}(cancel,t_{3})=x(y_b,t_3)+ \frac{Q}{e^{r_b\tau_b}}-f_b$ and 
$u_{A,int}(cancel,t_{3})=\frac{x_a}{e^{r_a(\tau_a+t_{\epsilon})}}+\frac{1.5Q}{e^{r_a (t_{\epsilon}+\tau_a) }}-2f_a$.
We observe that it is better to cancel the swap than wait for the contract to expire as \textbf{B} will lose his griefing-premium in the process. Thus \emph{cancel} strictly dominates \emph{stop} in our protocol. \textbf{B} will continue at $t_3$ over canceling the swap, if the following condition holds: $u_{B,int}(cont,t_{3})>u_{B,int}(cancel,t_{3})$

At time $t_2$, \textbf{A} decides to continue, then utility is:
$u_{A,int}(cont,t_{2})=  \theta_2\Big (\frac{\bigint_{x^3_1}^{x^3_2}  P(p,x(y,t_2),\tau_a) u_{A,int}(cont,t_{3}) \,dp}{e^{r_a\tau_a}} +
\frac{\bigint_{x^3_2}^{\infty}  P(p,x(y_b,t_2),\tau_a) u_{A,int}(stop,t_{3}) \,dp}{e^{r_a\tau_a}}\Big)+(1-\theta_2)\Big(\frac{u_{A,int}(stop,t_3)}{e^{r_a\tau_a}}\Big)$ and $u_{B,int}(cont,t_{2})=  \Big (\frac{\bigint_{x^3_1}^{x^3_2}  P(p,x(y,t_2),\tau_a) u_{B,int}(cont,t_{3}) \,dp}{e^{r_b \tau_a}} + \\
\frac{\bigint_{x^3_2}^{\infty}  P(p,x(y_b,t_2),\tau_a) u_{B,int}(stop,t_{3}) \,dp}{e^{r_b \tau_a}}\Big)$
At time $t_2$, \textbf{A} decides to abort then \textbf{B} initiates cancellation by releasing preimage of $H_2$. Note that \textbf{A} will not take any action if she intends to cancel as she has not locked any coins. Thus both cancel and stop means the same payoff for \textbf{A}. The payoff is defined as  $u_{A,int}(stop,t_{2})=u_{A,int}(cancel,t_{2})=x_a+1.5Q$ and $u_{B,int}(cancel,t_{2})=x(y_b,t_2)+ \frac{Q}{e^{r_b(\tau_a+\tau_b)}}-f_b$.
\textbf{A} will continue at $t_2$ over stopping the swap, if the following condition holds: $u_{A,int}(cont,t_{2})>u_{A,int}(cancel,t_{2})$.
\begin{proposition}
Quick Swap is more participant-friendly compared to HTLC-based atomic swap.
\end{proposition}
\emph{Proof}: In \emph{Quick Swap}, for given values of $\theta_1$ and $\theta_2$, success rate or $SR$ is function of $x_a$ (or \textbf{A}'s tokens) since there is no delay involved. It is expressed as: 

\begin{equation}
\begin{matrix}
    SR(x_a)=  \bigints_{x^3_1[x_a]}^{x^3_2[x_a]} A(x_a) B(x_a) \,dp 
    \end{matrix}
\end{equation}
where $A(x_a)= \theta_2 P(p,x(y_b,t_2),\tau_a)$ and $B(x_a)=  \theta_1\Big(1-C(x(y,t_4)[x_a],p,\tau_b)\Big) $

\textbf{A} is able to estimate the success rate now as it is dependent solely on $x_a$. There is no uncertainty involved, unlike in HTLC-based atomic swap where a higher delay leads to violation of participation constraint. Additionally, the range of $x_a$ for which \emph{Quick Swap} has a non-zero success rate is larger. The provision of cancellation, even before the elapse of the contract's locktime, makes the protocol robust and more resilient to fluctuation in asset price.
\subsection{Discussions}
(a) Our protocol requires an extra round compared to \emph{HTLC}-based atomic swap. The number of transactions created for \emph{Quick Swap} is double the number of transactions needed for the latter.\\
(b) The parameter $D$ can be chosen by $\textbf{A}$ and $\textbf{B}$, after they have negotiated with each other. A rational honest party will choose a shorter delay for keeping the griefing-premium locked.

\section{Construction of a fair multiparty cyclic atomic swap}

Suppose \emph{Alice} wants to exchange coins with \emph{Bob} where the former has some Bitcoins and later has Ethers, but \emph{Bob} wants to exchange his Ethers for Litecoins. In such a scenario, they take the help of some intermediaries for assisting in the exchange of coins. If we consider just a three-party situation, then there may exist a participant \emph{Carol} who is willing to exchange Litecoins for Bitcoins. So \emph{Alice} send $x_a$ BTC to \emph{Carol} and \emph{Carol} sends $z_c$ LTC to \emph{Bob}, and finally \emph{Bob} sends $y_b$ ETH to \emph{Alice}. In a real situation, \emph{Carol} will charge a fee from \emph{Alice} for facilitating the swap but we ignore the fee in this paper. 

\emph{Problem of Griefing}: In an HTLC-based setting, \emph{Alice} samples a secret $s$, shares $H=\mathcal{H}(s)$ with \emph{Carol} and \emph{Bob}. \emph{Alice} forwards HTLC to \emph{Carol} locking $x_a$ BTC in \texttt{Chain-a} for $T_1$ unit of time contingent to providing preimage of $H$. The confirmation time of a transaction in \texttt{Chain-a} in $\tau_a$. \emph{Carol} forwards an HTLC to \emph{Bob} using the same condition for a time period of $T_2$ units where $T_2<T_1$, locking $z_c$ LTC in \texttt{Chain-b}. The confirmation time of a transaction in \texttt{Chain-b} in $\tau_b$, Finally, \emph{Bob} forwards the HTLC, locking $y_b$ BTC in \texttt{Chain-c} for time period of $T_3$ units where $T_3<T_2$. The confirmation time of a transaction in \texttt{Chain-c} in $\tau_c$. The problem of griefing persists as \emph{Carol} may not choose to lock coins based on the fluctuation rate of Bitcoin and Litecoin, and even if \emph{Carol} locks her coins, \emph{Bob} may abort. If all the parties have locked coins, \emph{Alice} may abort, and makes \emph{Bob} and \emph{Carol} suffer. We discuss a fix to this problem by extending \emph{Quick Swap} from a two-party setting to a three-party setting.

\begin{figure}[!bht]
\hspace*{-2cm}
    \includegraphics[scale=0.4]{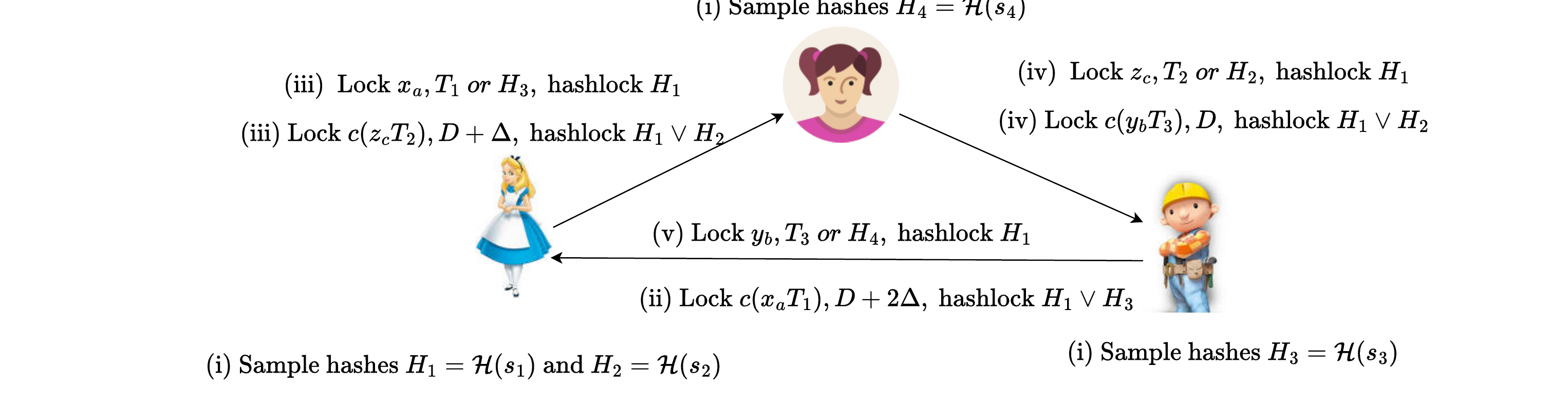}
    \caption{Quick Swap extended to three-party cyclic atomic swap}
    \label{figcyclic}
\end{figure}

\subsubsection*{High Level Overview of the protocol} We discuss the steps followed upon extending \emph{Quick Swap} to three party setting. The steps have been illustrated in Fig. \ref{figcyclic}.
\begin{itemize}
\item[(i)] Alice \emph{samples} hashes $H_1$ and $H_2$ using a randomly sampled secret $s_1$ and  $s_2$ respectively, and shares it with \emph{Bob} and \emph{Carol}. \emph{Bob} samples hash $H_3$ using secret $s_3$ and \emph{Carol} samples hash $H_4$ using secret $s_4$.
\item[(ii)] \emph{Bob} locks griefing-premium $c(x_aT_1)$ in \texttt{Chain-c} for a time-period of $D+2\Delta<T_3$ using hashlock $H_3 \vee H_1$, at time $t_1$.
\item[(iii)] \emph{Alice} locks principal amount $x_a$ for time period $T_1$ using hashlock $H_1$  at time $t_2=t_1+\tau_a$, with the provision of refunding earlier if preimage of $H_3$ is revealed. She samples a hash $H_2$ using secret $s_2$, and locks griefing-premium $c(z_cT_2)$ for a time period $D+\Delta$, hashlock $H_1\vee H_2$ in \texttt{Chain-a} at time $t_2$.
\item[(iv)] \emph{Carol} locks principal amount $z_c$ for time period $T_2$, hashlock $H_1$,  at time $t_3=t_2+\tau_b$ in \texttt{Chain-b}. She can refund before $T_2$ elapses if preimge of $H_2$ is revealed. \emph{Carol} samples hash $H_4$ using secret $s_4$, and locks griefing-premium $c(y_bT_3)$ for timperiod $D$, hashlock $H_1\vee H_4$ in \texttt{Chain-b}. 
\item[(v)] Finally, \emph{Bob} locks $z_c$ coins in \texttt{Chain-c} for timeperiod $T_3$, hashlock $H_1$, at time $t_2=t_1+\tau_a$. He has a provision to refund earlier if the preimage of $H_4$ is revealed. 
\end{itemize} 
To initiate the swap, \emph{Alice} reveals secret $s_1$ and everyone is able to unlock their griefing-premium and the swapped coins. If any of the parties want to cancel, he or she will choose to reveal either secret $s_2,s_3$ or $s_4$.
\subsection{Generic n-party fair cyclic atomic swap}
\subsubsection{System Model \& Assumption}
Party $P_0$ wants to exchange $a_0$ coins for $a_n$ coins of party $P_n$, taking help of $n-1$ intermediaries $P_1,P_2,\ldots,P_{n-1}$. A party $P_i$ has account in \texttt{Chain-i} and \texttt{Chain-($(i-1) \mod n+1)$ }. Blockchain \texttt{Chain-i} has transaction confirmation time $\tau_i$ where $i \in [0,n]$.
\subsubsection{Detailed Construction} We describe the steps:
\begin{itemize}
\item $P_0$ samples payment hash $\bar{H}$ and shares with neighbors $P_n$ and $P_1$.\\
    \item $P_n$ initiates the \emph{Locking Phase}, samples cancellation hash $H_n$. He locks griefing-premium $c(a_0T_0)$ for locktime $D+n\Delta<T_n$ in \texttt{Chain-n}, using hashlock $\bar{H} \vee H_n$, at time $t_1$.  \\
    \item Rest of the parties $P_i, i \in [0,n-1]$ does the following: $P_i$ generates a cancellation hash $H_i$ and locks $a_i$ coins for locktime $T_i$, using hashlock $\bar{H}$, at time $t_{i+2}=t_{i+1}+\tau_{(i-1) \mod n+1}$ in \texttt{Chain-i}. The coins can be refunded before $T_i$ if preimage of $H_{(i-1) \mod (n+1)}$ is revealed. He also locks griefing-premium $c(a_{i+1 }T_{i+1})$ at $t_{i+2}$, for locktime $D+(n-1-i)\Delta$, using hashlock $\bar{H} \vee H_{i}$. \\
  \item Finally, $P_n$ locks $a_n$ coins for locktime $T_n$, using hashlock $\bar{H}$, at time $t_{n+2}=t_{n+1}+\tau_{n-1}$ in \texttt{Chain-n}. He has an option to refund the coins if $P_{n-1}$ cancels the swap by revealing the preimage of $H_{n-1}$. The locktimes assigned follow a strictly decreasing order: $T_0>T_1>\ldots>T_n$.

\end{itemize}
\section{Related Works}
The HTLC-based atomic swap was first proposed in \cite{TN13}. However, the design lacks fairness and is susceptible to griefing attacks. Later Hao et al. \cite{han2019optionality} suggested the use of premium to counter griefing attacks. However, the protocol assumed that in a two-party setting where \emph{Alice} wants to exchange currency with \emph{Bob}, only \emph{Alice} can be at fault. So she must lock premium and \emph{Bob} is not required to do so. In an American-style option-based swap, \emph{Bob} gets the premium even though \emph{Alice} initiates the swap on time. In currency exchange-based atomic swap, \emph{Bob} gets the premium if \emph{Alice} doesn't respond within the time period of the contract. The protocol is not fair as \emph{Bob} can grief as well. The construction cannot be realized in Bitcoin scripts as it requires the inclusion of an additional opcode. 

Similar work has been done that talks about locking premium by both the parties involved in exchanging currency \cite{xue2021hedging}. However, the protocol is not compatible with Bitcoin scripts and suffers from the problem of mismatched premiums, and lacking fairness. Further, the authors have bootstrapped the premium, whereby small valued premiums get locked first, and with each iteration, the premium amount increases. This leads to multiple round communication, creation of multiple contracts for each iteration, and longer lock time than \cite{TN13} and griefing on the locked-up premium is possible \cite{9805490}. Nadahalli et al. \cite{9805490} have proposed a protocol that is \emph{grief-free} and compatible with Bitcoin scripts. The protocol is efficient regarding the number of transactions and
the worst-case timelock for which funds remain locked. However, the problem of mismatched premium exists. The model lacks flexibility due to the coupling of premium with the principal amount, and thus cannot be extended to multi-party atomic swap setting involving more than two blockchains \cite{H18}. Our proposed protocol overcomes several such shortcomings. However, there is a constant factor increase in overhead of transaction and communication round compared to \cite{han2019optionality} and \cite{9805490}. We summarize the discussion by performing a comparative analysis of \emph{Quick Swap} with other state-of-the-art protocols in Table \ref{tab:my_label}.
\begin{table}[ht]
    \centering
\scalebox{0.6}{
    \begin{tabular}{|c|c|c|c|c|c|c|}
    \hline
          &TN \cite{TN13} &HLY    \cite{han2019optionality} &XH  \cite{xue2021hedging} &NKW \cite{9805490}  &\emph{Quick}\\
          &Swap &Swap &Swap &Swap  &\emph{Swap}\\
    \hline

	$P1$  &\xmark  &Partial  &\cmark &\cmark &\cmark\\
			$P2$ &\xmark &\xmark  &\xmark &Partial &\cmark\\
			$P3$ &\xmark &\xmark  &\xmark &Partial &\cmark\\
			$P4$ &NA &\xmark  &\xmark &\xmark &\cmark\\
			$P5$ &\cmark &\xmark  &\xmark &\cmark &\cmark\\
          $P6$ &NA &\xmark  &\xmark &\xmark &\cmark\\
          
    \hline
    \end{tabular}}
    \caption{Comparative Analysis of \emph{Quick Swap} with existing Atomic Swap protocols in terms of $P1$: Countering griefing attack, $P2$: Cancellation Allowed, $P_3$ Counters speculation, $P4$ Fairness of premium locked, $P_5$ Supported by Bitcoin script and $P6$: Extension to multi-party cyclic swap}
    
    \label{tab:my_label}
\end{table}




\section{Conclusion}
In this paper, we perform a game-theoretic analysis of HTLC-based atomic swap. We observe that the protocol lacks fairness and it is not at all participant-friendly. We propose \emph{Quick Swap} that is robust and allows faster settlement of the transaction. We discuss the step for extending \emph{Quick Swap} to a multiparty setting involving more than two blockchains. As a part of our future work, we would like to analyze \emph{Quick Swap} in presence of rational miners in underlying Blockchains. 
\section*{Acknowledgement}
We thank Dr. Abhinandan Sinha, Assistant Professor at Ahmedabad University, for his initial valuable comments on this work. 
\bibliographystyle{ACM-Reference-Format}
\bibliography{SWAP}

\end{document}